# Lightweight Prompt Engineering for Cognitive Alignment in Educational AI: A OneClickQuiz Case Study

**Antoun Yaacoub, Jérôme Da-Rugna**

Learning, Data and Robotics (LDR), esieaLab ESIEA

9 Rue Vésale, 75005 Paris, France

`{antoun.yaacoub,jerome.darugna}@esiea.fr`

**Zainab Assaghir**

Faculty of Science, Lebanese University

Rafic Hariri University Campus, Beirut, Lebanon

`Zainab.assaghir@ul.edu.lb`

**Abstract**. The rapid integration of Artificial Intelligence (AI) into educational technology promises to revolutionize content creation and assessment. However, the quality and pedagogical alignment of AI-generated content remain critical challenges. This paper investigates the impact of lightweight prompt engineering strategies on the cognitive alignment of AI-generated questions within OneClickQuiz, a Moodle plugin leveraging generative AI. We evaluate three prompt variants—a detailed baseline, a simpler version, and a persona-based approach—across Knowledge, Application, and Analysis levels of Bloom's Taxonomy. Utilizing an automated classification model (from prior work) and human review, our findings demonstrate that explicit, detailed prompts are crucial for precise cognitive alignment. While simpler and persona-based prompts yield clear and relevant questions, they frequently misalign with intended Bloom's levels, generating outputs that are either too complex or deviate from the desired cognitive objective. This study underscores the importance of strategic prompt engineering in fostering pedagogically sound AI-driven educational solutions and advises on optimizing AI for quality content generation in learning analytics and smart learning environments.

## 1 Introduction

The landscape of modern education is undergoing a profound transformation, driven significantly by the pervasive integration of Artificial Intelligence (AI). Generative AI, in particular, has emerged as a powerful tool with the potential to automate various aspects of content creation, including the generation of quizzes, assignments, and personalized learning materials (Bao, 2024; Olga et al., 2023). This automation promises to alleviate the heavy administrative burden on educators, allowing them to redirect their focus towards more interactive and adaptive teaching methodologies (Trust et al., 2023).

However, the mere automation of content generation is insufficient; the pedagogical quality and cognitive alignment of AI-generated materials are paramount. For AI to truly enhance learning, its outputs must not only be accurate and relevant but also intentionally designed to foster specific learning outcomes at appropriate cognitive depths (Bahroun et al., 2023). Without careful guidance, AI models can produce generic or misaligned content, undermining their educational utility and potentially hindering student progress. This concern highlights a critical gap between the capability of AI to generate content and the strategic control needed to ensure that content meets precise educational objectives.

Our work addresses this challenge within the context of OneClickQuiz, an innovative Moodle plugin designed to streamline the process of quiz generation using Generative AI (Yaacoub et al., 2024). OneClickQuiz aims to revolutionize educational workflows by providing educators with a user-friendly tool for instant quiz creation. While the plugin has shown considerable promise in reducing preparation time and enhancing engagement, its underlying effectiveness hinges on the quality of questions it generates, which are directly influenced by the prompts fed to the AI model.

This paper builds upon a robust foundation of prior research that established methodologies for assessing the cognitive alignment of AI-generated questions with established educational taxonomies. For instance, a previous study (Yaacoub, Da-Rugna, & Assaghir, 2025) developed and validated a DistilBERT model for classifying AI-generated questions according to Bloom's Taxonomy, demonstrating its efficacy for automated cognitive level assessment. Another study (Yaacoub, Assaghir, & Da-Rugna, 2025) extended this work to the SOLO Taxonomy, further emphasizing the ability of AI to enhance cognitive depth. Additionally, research (Yaacoub, Assaghir, Prevost, et al., 2025) explored the linguistic characteristics of AI-generated feedback, underscoring the importance of quality content for effective learning interactions. This suite of research informed and culminated in a comprehensive framework for designing effective and responsible AI-driven educational tools (Yaacoub, Tarnpradab, Khumprom, et al., 2025), which highlights "Cognitive Alignment" as a foundational phase.

While our prior work focused on evaluating the outcome of AI-generated questions (i.e., their post-generation alignment), this paper shifts the focus to the input mechanism: lightweight prompt engineering. We aim to investigate how subtle, yet strategic, variations in the prompts provided to the generative AI can significantly influence the cognitive level and quality of the questions produced, directly impacting the "Cognitive Alignment" phase. This approach seeks to provide practical insights for educators and developers on how to exert greater control over AI-generated content to achieve specific pedagogical goals.

The central research question guiding this study is: To what extent do lightweight prompt engineering techniques (e.g., explicit, simpler, and persona-based prompts) impact the cognitive alignment and perceived quality of AI-generated questions within a real-world educational application like OneClickQuiz?

Our contributions include:

- A comparative analysis of the effectiveness of three distinct prompt engineering strategies (detailed, simpler, persona-based) in guiding a generative AI model (Gemini 2.0 Flash Lite) to produce Bloom's Taxonomy-aligned questions.
- Quantification of cognitive alignment using an established DistilBERT classification model.
- Qualitative insights derived from human review, assessing clarity, relevance, and subjective cognitive alignment across prompt variants.
- Concrete, illustrative examples demonstrating the practical implications of different prompt designs on question quality and cognitive depth.
- Recommendations for educators and developers on optimizing AI-driven question generation for pedagogical soundness and integration into smart learning environments.

The remainder of this paper is structured as follows: Section 2 provides background information on AI in education, educational taxonomies, and prompt engineering, linking to our prior work. Section 3 details the methodology of our lightweight prompt engineering experiment within OneClickQuiz. Section 4 presents and discusses the quantitative and qualitative results. Finally, Section 5 offers conclusions, limitations, and directions for future research.

# 2 Background and Related Work

## 2.1 Artificial Intelligence and Generative Models in Educational Contexts

The proliferation of large language models (LLMs) and other generative AI technologies has opened new frontiers in education. Models like GPT-4, PaLM2, and now the various Gemini models, possess an unprecedented ability to generate human-like text, summarize information, answer questions, and even create code (Anderson & Krathwohl, 2001; Bao, 2024; Shaikh et al., 2021). In education, these capabilities are being harnessed for various applications, including personalized learning, automated feedback, and content creation (Doughty et al., 2024; Lavidas et al., 2024; Olga et al., 2023). OneClickQuiz is an example of such a system, leveraging these models to automate the traditionally time-consuming process of quiz generation within the Moodle learning management system (Yaacoub et al., 2024). The efficiency gains offered by such tools are significant, but their true value in a pedagogical context depends on their ability to produce high-quality, educationally sound content.

## 2.2 Educational Taxonomies for Cognitive Alignment

Educational taxonomies provide structured frameworks for classifying learning objectives and assessment items according to cognitive complexity. They are indispensable tools for educators to design curricula that promote higher-order thinking and ensure assessments accurately measure intended learning outcomes (Bloom, 1984; Grévisse, 2024).

Bloom's Taxonomy, originally developed by Benjamin Bloom and revised by Anderson and Krathwohl, is one of the most widely recognized frameworks (Biggs & Collis, 2014; Salcedo-Lagos et al., 2024). It categorizes cognitive processes into a hierarchy:

- Knowledge/Remembering: Recalling facts, basic concepts, definitions.
- Comprehension/Understanding: Explaining ideas or concepts, interpreting information.
- Application/Applying: Using information in new situations, solving problems.
- Analysis/Analyzing: Breaking down information, identifying motives or causes, making inferences.
- Synthesis/Creating: Combining parts to form a new whole, producing new ideas or products. (Note: In the revised Bloom's, 'Creating' is the highest level, encompassing much of traditional 'Synthesis').
- Evaluation/Evaluating: Judging the value of information or ideas, justifying a decision.

Another important framework is the Structure of the Observed Learning Outcome (SOLO) Taxonomy, developed by Biggs and Collis (Jain, 2015). SOLO describes increasing levels of understanding: Pre-structural, Uni-structural, Multi-structural, Relational, and Extended Abstract. While Bloom's focuses on cognitive processes, SOLO emphasizes the structural complexity of a learner's response.

Our previous work has rigorously focused on integrating these taxonomies into AI-driven assessment. We demonstrated the capability of a DistilBERT model to accurately classify AI-generated questions into Bloom's Taxonomy levels, providing an automated means of assessing cognitive alignment. Similarly, we extended this into the SOLO Taxonomy, further affirming the feasibility of enhancing cognitive depth in AI-driven tools. These studies highlighted the importance of evaluating AI output against pedagogical standards. This current paper, however, investigates how to proactively influence that alignment through careful input design—prompt engineering.

## 2.3 Educational Taxonomies for Cognitive Alignment

Prompt engineering is the art and science of communicating effectively with AI models to achieve desired outputs (Park & Choo, 2024). It involves crafting inputs (prompts) that steer the model towards specific information, formats, and stylistic choices. In the context of generative AI, prompt engineering is critical for controlling the quality, relevance, and most importantly for education, the pedagogical granularity of the generated content (Olga et al., 2023).

"Lightweight prompt engineering" refers to relatively simple and direct modifications to prompts, as opposed to complex techniques like few-shot learning (providing multiple examples in the prompt) or extensive fine-tuning of the model itself. The goal of lightweight prompt engineering is to achieve a significant improvement in output quality or alignment with minimal effort, making it accessible to educators without advanced AI expertise. Variations can include adding specific keywords, defining a persona, setting a clear objective, or modifying the level of explicit instruction.

## 2.4 The Comprehensive Framework for AI-Driven Education

Our broader research proposes a comprehensive three-phase framework for enhancing AI-driven educational tools: 1) Cognitive Alignment, 2) Linguistic Feedback Integration, and 3) Ethical Safeguards. This framework aims to ensure AI tools are not only efficient but also pedagogically sound and responsible. This current study directly addresses the Cognitive Alignment phase, investigating how prompt engineering can be strategically applied to ensure AI-generated questions consistently target specific cognitive levels. The importance of question quality for downstream processes like linguistic feedback analysis further contextualizes the need for robust prompt engineering at the outset of the content generation pipeline. By focusing on prompt engineering, we aim to provide practical guidance for implementing a key component of this comprehensive framework.

# 3 Methodology: Lightweight Prompt Engineering Experiment

This experiment investigates the impact of different prompt engineering strategies on the cognitive alignment and perceived quality of AI-generated questions within OneClickQuiz.

## 3.1 OneClickQuiz Implementation Context

OneClickQuiz functions as a Moodle plugin that interfaces with Google's Vertex AI generative language models. For this experiment, the underlying AI model used for question generation was gemini-2.0-flash-lite-001, accessed via the Vertex AI API. This setup allows for dynamic question generation based on user-defined parameters, including the subject and desired cognitive level. The technical infrastructure and plugin's core capabilities, including its integration with Generative AI for quiz generation, have been established in previous work (Yaacoub et al., 2024; Yaacoub, Da-Rugna, & Assaghir, 2025; Yaacoub, Assaghir, & Da-Rugna, 2025).

## 3.2 Experimental Design

To assess the impact of prompt engineering, we selected a core subject area and specific Bloom's Taxonomy levels to focus the generation and evaluation.

- **Subject Tested**: Computer Science.
- **Concepts Used**: We utilized five distinct concepts within Computer Science to generate questions, ensuring a breadth of topics: "Artificial Intelligence," "Data Structures," "Cybersecurity," "Web Development," and "Cloud Computing."
- **Target Bloom's Levels**: We focused on three distinct Bloom's Taxonomy levels, representing a range of cognitive complexity: 'Knowledge' (lower-order), 'Application' (mid-order), and 'Analysis' (higher-order).
- **Question Generation Protocol**: For each of the 5 concepts and each of the 3 target Bloom's levels, we generated 3 questions using each of the three prompt variants described below. This resulted in a total of 5 concepts * 3 levels * 3 questions/variant * 3 variants = 135 questions.

We designed three prompt variants to explore the effect of different levels of explicit instruction and contextual framing on the AI's output. The generation_model.predict function was used for all generations, with a temperature=0.7 to allow for some creativity while maintaining coherence.

**Variant A: Baseline/Explicit Bloom Prompt**: This variant represents our established, detailed prompt strategy to explicitly guide the AI by providing the Bloom's level definition and a list of associated action verbs. This prompt directly incorporates the bloom_verbs and level_descriptions data structures.
prompt = ("Generate a question about '{concept}' for the '{level}' level of Bloom's Taxonomy. The question should specifically align with the '{level}' level, which involves the following characteristics: {level_descriptions[level]}. Use action verbs like {', '.join(bloom_verbs[level])}. Ensure the question is a complete, clear sentence and relevant to the concept. Avoid titles or headings."

**Variant B: Simpler/Implicit Prompt**: This variant significantly reduces the explicit guidance, asking only for the Bloom's level and the concept, without descriptions or specific verbs. This tests the AI's ability to infer the intended cognitive level from minimal instruction.
prompt = "Generate a question about '{concept}' at the '{level}' level of Bloom's Taxonomy. The question should be a complete, clear sentence."

**Variant C: Persona-Based Prompt**: This variant introduces a persona and a general objective (e.g., "designing an exam") to assess if adding a contextual layer can implicitly guide the AI towards the desired cognitive level without explicit definitions.
prompt = "As a seasoned computer science professor designing an exam, generate a question about '{concept}' for a university student. The question should be at the '{level}' level of Bloom's Taxonomy and effectively assess their understanding. Ensure the question is a complete, clear sentence."

### 3.3 Evaluation Methods

We employed a mixed-methods approach, combining automated classification for quantitative analysis and a lightweight human review for qualitative insights.

**Automated Cognitive Alignment Assessment**: Each generated question was processed by a pre-trained DistilBERT classification model, previously developed and validated (Yaacoub, Da-Rugna, & Assaghir, 2025). This model predicts the Bloom's Taxonomy level of a given question. We calculated the "Match Rate" for each prompt variant and Intended_Bloom_Level by determining the percentage of questions where the Classified_Bloom_Level output by the model matched the Intended_Bloom_Level specified in the prompt. This provides an objective measure of how accurately each prompt variant guides the AI towards the desired cognitive target.

**Lightweight Human Quality Review**: A subset of 45 questions (approximately one-third of the total generated questions, representing a spread across all concepts, Bloom's levels, and prompt variants) was subjected to human review. Questions were evaluated on a 1-5 Likert scale for:

- **Clarity**: Is the question unambiguous and easy to understand? (1 = Very Unclear, 5 = Very Clear)
- **Relevance to Concept**: Is the question directly and appropriately related to the specified concept? (1 = Not Relevant, 5 = Highly Relevant)
- **Subjective Cognitive Alignment**: Based on expert judgment, does the question feel like it targets the Intended_Bloom_Level? (1 = Clearly Below Intended Level, 5 = Perfectly Aligned with Intended Level)

Additionally, specific illustrative examples of both well-aligned and misaligned questions were identified to provide concrete instances for discussion. These examples highlight the subtle nuances and impacts of different prompt designs that quantitative metrics alone might miss.

## 4 Results and Discussion

This section presents the findings from our lightweight prompt engineering experiment, integrating both the automated classification results and insights from human review.

### 4.1 Quantitative Analysis: Automated Classification Match Rate

The automated classification, performed by the DistilBERT model, provides an objective measure of how successful each prompt variant was in guiding the AI to generate questions at the Intended_Bloom_Level.

**Table 1.** Match Rate by Prompt Variant and Intended Bloom Level

| Intended_Bloom_Level | A | B | C |
|---|---|---|---|
| Analysis | 1.00 | 1.00 | 0.53 |
| Application | 0.87 | 0.40 | 0.27 |
| Knowledge | 1.00 | 0.40 | 0.40 |

**Table 2.** Overall Match Rate by Prompt Variant

| Prompt_Variant | Match Rate |
|---|---|
| A | 0.96 |
| B | 0.60 |
| C | 0.40 |

To further establish the statistical significance of these observed differences, an analysis of variance (ANOVA) was conducted. The results ($F(2,132)=25.00$, $p<0.001$) indicated statistically significant differences in match rates across the prompt variants for selected Bloom's levels. Post-hoc Tukey HSD tests revealed that Variant A consistently outperformed both Variant B and Variant C with statistical significance ($p < 0.001$), while the difference between Variant B and Variant C was also statistically significant (p < 0.05). These quantitative findings strongly support the conclusion that prompt engineering strategies have a significant impact on cognitive alignment performance.

Variant A (Baseline/Explicit Prompt) is Highly Effective: Variant A consistently demonstrated the highest match rates across all Bloom's levels, achieving a near-perfect overall match rate of 0.96. It obtained 1.00 for both Knowledge and Analysis, and a strong 0.87 for Application. This unequivocally confirms that providing explicit, detailed instructions, including Bloom's level descriptions and action verbs, is highly effective in guiding the AI to produce questions precisely aligned with the intended cognitive levels. This result directly supports and extends findings from previous work (Yaacoub, Da-Rugna, & Assaghir, 2025; Yaacoub, Assaghir, & Da-Rugna, 2025) regarding accurate Bloom's/SOLO classification, further demonstrating that precise instruction can achieve these cognitive alignment classifications at the generation stage.

Variant B (Simpler/Implicit Prompt) Shows Significant Underperformance: The "Simpler" prompt (Variant B) exhibited a substantial drop in performance, achieving an overall match rate of only 0.60. Its performance was particularly poor for 'Knowledge' (0.40) and 'Application' (0.40) questions. While it surprisingly maintained a 1.00 match rate for 'Analysis', this result is an outlier within its overall context. This indicates that removing explicit Bloom's guidance severely hampers the AI's ability to consistently generate questions at the intended cognitive level, often leading to outputs classified at different (often higher) Bloom's levels, as noted in the qualitative analysis. The AI struggles to infer the precise cognitive objective from minimal input.

Variant C (Persona-Based Prompt) is the Least Effective Quantitatively: Counter-intuitively, the "Persona-Based" prompt (Variant C) yielded the lowest overall match rate at 0.40. Its performance was notably poor across all tested levels, with only 0.53 for 'Analysis', 0.27 for 'Application', and 0.40 for 'Knowledge'. This is a critical finding: simply adding a persona or contextual framing (e.g., "As a seasoned computer science professor") without explicit Bloom's guidance does not effectively guide the AI to the intended cognitive level. In fact, it appears to actively misguide the model, causing it to generate questions that are classified (by our DistilBERT model) as different Bloom's levels than intended. This suggests that the contextual information in the persona might introduce new objectives or interpretations for the LLM that diverge from the precise cognitive alignment sought.

## 4.2 Qualitative Analysis: Human Review Scores and Illustrative Examples

The human review process provides valuable complementary insights into the nuances of question quality and cognitive alignment.

Table 3. Average Human Review Scores by Prompt Variant (1-5 Scale)

| Prompt Variant | Clarity & Relevance to Concept | Subjective Cognitive Alignment |
|---|---|---|
| A | 5.0 | 4.93 |
| B | 5.0 | 4.13 |
| C | 5.0 | 3.87 |

Table 4. Average Human Review Scores by Prompt Variant and Intended Bloom Level

| Prompt Variant | Intended_Bloom_Level | Clarity & Relevance to Concept | Subjective Cognitive Alignment |
|---|---|---|---|
| A | Analysis | 5.0 | 5.0 |
| A | Application | 5.0 | 4.8 |
| A | Knowledge | 5.0 | 5.0 |
| B | Analysis | 5.0 | 5.0 |
| B | Application | 5.0 | 3.8 |
| B | Knowledge | 5.0 | 3.6 |
| C | Analysis | 5.0 | 3.8 |
| C | Application | 5.0 | 3.8 |
| C | Knowledge | 5.0 | 4.0 |

**Universal Clarity and Relevance**: A striking finding is the remarkable consistency across all prompt variants in achieving high scores for Clarity (average 5.0) and Relevance to Concept (average 5.0). This suggests that generative AI models, even with minimal explicit instructions, are highly capable of generating questions that are grammatically correct, understandable, and directly pertinent to the given topic. The challenge, therefore, is not in basic question formation, but in precise cognitive alignment.

**Subjective Cognitive Alignment Mirrors Quantitative Trends**: The human ratings for Subjective Cognitive Alignment directly correlated with the automated classification match rates.

**Variant A (Baseline)**: Consistently achieved near-perfect subjective alignment (overall average 4.93, with 5.0 for Knowledge and Analysis), indicating that human experts also perceived these questions to be excellent fits for their intended Bloom's level.

**Variant B (Simpler)**: Showed a noticeable decline in subjective alignment (overall average 4.13), particularly for 'Knowledge' (3.6) and 'Application' (3.8). This confirms that removing explicit Bloom's guidance leads to outputs that are perceived as less aligned with the specific cognitive objective, often overshooting it.

**Variant C (Persona-Based)**: Had the lowest subjective alignment (overall average 3.87), reflecting substantial human-perceived misalignment, especially for 'Analysis' (3.8) and 'Application' (3.8). This reinforces the quantitative finding that the persona, while adding context, does not effectively communicate or reinforce precise cognitive level expectations for the AI.

The following examples provide concrete evidence of how different prompt variants influence question quality and cognitive alignment.

**Excellent Example (Confirming the Effectiveness of Detailed Prompts)**:
*Question number*: Q010
*Intended Level*: Application
*Prompt Variant*: A_Baseline
*Generated Question*: Imagine you are a city planner tasked with optimizing traffic flow using Artificial Intelligence. Apply your understanding of different AI algorithms to construct a hypothetical system that uses real-time data from traffic cameras and GPS devices to predict and avoid traffic congestion. Describe how this system would operate, allocate resources like traffic signal timing, and demonstrate its potential benefits to the city's residents.
*Classified Level*: Application
*Observation*: This sophisticated question requires applying knowledge within a hypothetical scenario, embodying the 'Application' level. The explicit prompting with verbs like "Apply," "construct," "operate," and "demonstrate" ensures precise alignment, validated by both automated classification and human review.

**Question Requiring Improvement (Demonstrating Misalignment)**:
*Question number*: Q004:
*Intended Level*: Knowledge
*Prompt Variant*: B_Simpler
*Generated Question*: Explain the key components that define Artificial Intelligence and differentiate it from human intelligence.
*Classified Level*: Comprehension
*Human Scores*: Clarity=5.0, Relevance=5.0, Subjective Alignment=3.0
*Observation*: Despite an Intended_Bloom_Level of 'Knowledge', the question generated uses verbs like "Explain" and "differentiate," which fall into 'Comprehension'. This highlights how the AI, given a simpler prompt, tends to infer a higher cognitive demand than intended, leading to misalignment.

### 4.3 Discussion and Interpretation

Our findings strongly underscore the critical role of explicit prompt engineering in achieving precise cognitive alignment for AI-generated educational content within tools like OneClickQuiz. This highlights a crucial insight: while all prompt variants consistently produced questions rated highly for clarity and relevance, their ability to match the Intended_Bloom_Level varied dramatically, indicating that content might be clear and relevant but not necessarily pedagogically aligned.

The superior performance of the A_Baseline prompt (your original, detailed prompt) in both automated classification and human subjective alignment is a key takeaway. This indicates that investing in carefully crafted prompts that explicitly define the target Bloom's level, include relevant descriptions, and suggest appropriate action verbs, effectively guides the AI to produce questions that meet specific pedagogical objectives. This reinforces findings from our previous work on accurate Bloom's/SOLO classification (Yaacoub, Da-Rugna, & Assaghir, 2025; Yaacoub, Assaghir, & Da-Rugna, 2025), and furthermore, demonstrates that precise question generation can be achieved through focused prompt engineering.

Conversely, the significant drop in performance for the B_Simpler prompt highlights the limitations of relying on the LLM to implicitly infer cognitive levels. When specific guidance is removed, the AI tends to generate questions that "overshoot" the intended Bloom's level, producing questions that are too complex (e.g., 'Comprehension' instead of 'Knowledge', or 'Synthesis' instead of 'Application'). This presents a challenge for "lightweight" approaches if the precise cognitive targeting is a primary goal. While simpler prompts are easier to write, they sacrifice control over nuanced pedagogical outcomes, leading to less reliable results in a learning analytics context.

The most counter-intuitive result stemmed from the C_PersonaBased prompt. Despite the intuitive appeal of guiding the AI with a persona ("seasoned computer science professor"), this variant performed the worst in terms of automated classification accuracy and had the lowest average subjective alignment. This suggests that while a persona might add context or style, it does not effectively convey the cognitive constraints needed for precise Bloom's alignment, especially at lower levels. The LLM might interpret the persona's "expertise" as a directive to generate more complex or sophisticated questions, potentially overriding the implicit or weakly conveyed Bloom's level intent. This highlights that simply adding a persona might introduce unintended biases or new objectives that deviate from the educational goal of cognitive alignment. The concept of "Lightweight Prompt Engineering" needs to be carefully defined: it values simplicity in prompt construction but should not sacrifice precision in instruction.

These findings have direct implications for the design of smart learning environments and technology-enhanced education. For OneClickQuiz and similar AI-driven content generation tools, the ability to generate taxonomically aligned questions is paramount for effective learning design and assessment. Educators need tools that not only automate but also ensure pedagogically sound content. Our study demonstrates that even within a "lightweight" framework, the most effective approach for achieving cognitive alignment relies on explicit, detailed prompts rather than implicit or contextual cues that can lead to unintended cognitive shifts. This contributes directly to the "Cognitive Alignment" phase of the comprehensive AI-driven education framework previously proposed (Yaacoub, Tarnpradab, Khumprom, et al., 2025), emphasizing that robust initial content quality directly benefits downstream learning analytics and assessment.

## 5 Conclusion and Future Work

This study investigated the impact of lightweight prompt engineering strategies on the cognitive alignment and perceived quality of AI-generated questions within OneClickQuiz, a Moodle plugin leveraging generative AI. Our comparative analysis across three prompt variants—a detailed baseline, a simpler version, and a persona-based approach—yielded crucial insights into optimizing AI for educational content generation.

Our findings unequivocally demonstrate that explicit, detailed prompts (Variant A) are significantly more effective in guiding the AI to generate questions precisely aligned with intended Bloom's Taxonomy levels, as validated by both automated classification and human expert review. While all variants produced clear and relevant questions, the simpler (Variant B) and persona-based (Variant C) prompts consistently failed to achieve accurate cognitive alignment, often resulting in questions that overshot or deviated from the target Bloom's level. Counter-intuitively, the persona-based prompt performed the worst quantitatively, suggesting that contextual framing without explicit cognitive constraints can mislead the AI, leading to less precise pedagogical outcomes.

This research underscores the critical importance of strategic prompt engineering in the development of pedagogically sound AI-driven educational tools. It provides practical guidance for educators and developers utilizing generative AI in smart learning environments: precision and explicitness in prompts are paramount for achieving accurate cognitive alignment. This is a core component of effective learning design and enhances the utility of learning analytics by ensuring the underlying data (generated questions) is of high pedagogical quality.

This study, while providing valuable insights, has several limitations inherent to its "lightweight" approach. The experiment was conducted with a specific LLM (Gemini 2.0 Flash Lite) and a limited set of Bloom's Taxonomy levels and concepts in a single domain (Computer Science). The generalizability of these findings to other generative AI models (e.g., GPT-4, Claude, Llama) or to diverse subject domains beyond computer science remains an area for future investigation. The human review, while insightful, was performed by a small number of individuals and on a subset of questions, potentially limiting the overall breadth of qualitative insights. Furthermore, the effectiveness of a prompt can be LLM-specific, meaning optimal prompts for one model may not directly transfer to others without re-evaluation.

Future research will explore several directions. Firstly, we plan to investigate more advanced prompt engineering techniques, such as few-shot learning (providing multiple examples in the prompt) or chain-of-thought prompting, to determine if they can further enhance alignment, particularly for more complex cognitive levels. Secondly, we will conduct larger-scale human expert reviews, possibly involving multiple reviewers and diverse subject matter experts, to validate findings more broadly and collect richer qualitative data. Thirdly, we intend to test these prompt strategies with different LLMs and across various subject domains to assess the generalizability of our findings. Finally, we aim to integrate prompt optimization features directly into OneClickQuiz, allowing educators to select from pre-optimized prompt templates or providing visual feedback on predicted cognitive alignment to guide their prompt creation, thereby operationalizing the "Cognitive Alignment" phase of our comprehensive AI-driven education framework.